\documentclass[12pt,preprint,a4]{aastex}  

\slugcomment{To appear in the Astrophysical Journal}

\begin{document}
\title{A Bayesian Approach to Solar Flare Prediction}
\author{M. S. Wheatland}
\affil{School of Physics, University of Sydney, NSW 2006, Australia}
\email{m.wheatland@physics.usyd.edu.au}

\begin{abstract}
A number of methods of flare prediction rely on classification of
physical characteristics of an active region, in particular optical
classification of sunspots, and historical rates of flaring for a 
given classification. However these methods largely ignore the number 
of flares the active region has already produced, in particular the 
number of small events. The past history of occurrence of 
flares (of all sizes) is an important indicator to future flare 
production. We present a Bayesian approach to flare prediction, 
which uses the flaring record of an active region together with 
phenomenological rules of flare statistics to refine an initial 
prediction for the occurrence of a big flare during a subsequent
period of time. The initial prediction is assumed to come from one 
of the extant methods of flare prediction. The theory of the method 
is outlined, and simulations are presented to show how the refinement 
step of the method works in practice.  
\end{abstract}


\section{Introduction}

Solar flares influence local `space weather,' and as a result there is 
a demand for accurate flare prediction. Unfortunately no reliable 
deterministic method of predicting a flare is known, and existing methods 
are probabilistic in nature. 

A number of methods discussed in the literature are based on a commonly
used white-light classification of sunspots, and the correlation 
between classification and flare occurrence. The McIntosh classification
(McIntosh 1990) categorizes a group of sunspots into one of 60 classes,
based on three parameters. Historical flare rates for each of the 
classifications were used by McIntosh (1990) as the basis of an 
`expert system' for flare prediction. The system, called Theophrastus
(the associated code is called THEO), also incorporates additional 
information including dynamical properties 
of spot growth, rotation and shear, magnetic topology inferred from 
sunspot structure, magnetic classification, and previous flare activity. 
The method is apparently somewhat subjective, involving rules of thumb 
incorporated by a human expert. A second approach using the McIntosh 
classification was presented by Bornmann and Shaw (1994). In this case 
multiple linear regression was used to determine the effective contribution 
of each of the McIntosh parameters to the rate of flaring, based on historical 
records of flaring. Codes based on the methods of McIntosh (1990) and 
Bornmann and Shaw (1994) are used by the Ionospheric Prediction 
Service (IPS) of Australia to issue flare predictions at their
Learmonth and Culgoora observatories.\footnote{See http://www.ips.gov.au.} 
Recently Gallagher, Moon and Wang (2002) implemented a system
using historical averages of flare numbers for McIntosh classifications
to predict a rate for an active region, and then converted this to
a probability of flaring in a day using the assumption of Poisson 
statistics. This prediction is given as part of the Big Bear Solar
Observatory Active Region Monitor (ARM).\footnote{See 
http://beauty.nascom.nasa.gov/arm/latest/.} Finally the US National
Oceanic and Atmospheric Administration (NOAA) issues flare 
probability forecasts for active regions
which include input from THEO.\footnote{See 
http://www.sec.noaa.gov/ftpdir/latest/daypre.txt.} 

A shortcoming of methods relying on correlations of flaring with 
active region classification based on historical records is that they
ignore the important information of how many flares the active region 
of interest has already produced. The system of McIntosh (1990) 
incorporates information about previous activity, but it is unclear 
how objectively this is done, and the information is limited to 
the number of large flares already produced by the given active region. 
In the flare prediction 
literature, the tendency of a region which has produced large flares in 
the past to produce large flares in the future is called persistence, 
which is recognised as one of the most reliable predictors for large 
flare occurrence in 24-hour forecasts (e.g.\ Neidig, Weiborg, \& 
Seagraves 1989). In this paper we argue that the history of occurrence of 
all flares (large and small) observed in a given active region is an 
important indicator as to how the region will flare in the future, and 
should be used in any prediction. A related criticism of methods based
on classification and historical records is that a given classification
may embrace active regions with a variety of flaring rates. If an
active region has a flaring rate differing from the average historical
rate for its class then the predictions will be in error. 

Studies of solar flare statistics provide simple phenomenological 
rules describing flare occurrence. It is well known that flares follow 
a power-law size distribution, where by size we mean e.g.\ peak flux 
in soft X-ray. More 
formally the flare frequency-size distribution $N(S)$ (i.e.\ the number 
of events per unit size $S$ and per unit time) may be written
\begin{equation}\label{eq:pldist}
N(S)=AS^{-\gamma}
\end{equation}
where $A$ and $\gamma$ are constants. The exact power-law
index $\gamma$ depends on the choice of the quantity $S$, but typically 
it is found to be in the range 1.5 to 2 (e.g.\ Crosby, Aschwanden,
\& Dennis 1992). The power law index $\gamma$ appears to be the same
in different active regions~\cite{whe00}, although there is some 
evidence that it varies with the solar cycle~\cite{bai93}. A second
simple rule concerns the way flares occur in time. Studies of the 
rate of occurrence of soft X-ray flares in individual active regions 
suggest that events occur as a Poisson process in time (e.g.\ Moon et 
al.\ 2001), although many active regions exhibit changes in the 
mean rate of events (Wheatland 2001).

In this paper we show how the observed record of flaring in an active
region may be used together with the phenomenological rules of 
flare statistics to objectively refine an initial flare prediction. 
The initial prediction may be based on the McIntosh classification, or 
may come from any other prediction method which does not consider the 
flare data. The new method
is envisaged to work as follows. When an active region appears at the 
east limb of the Sun, the best guess as to its future flare productivity 
comes from one of the conventional prediction methods. However, as the 
active region produces flares, the observed flare statistics are used to 
adjust the prediction for future flaring. After many flares have been
observed, the prediction for future flaring may be dominated by the 
contribution from the observed data. This process --- refining a 
probability estimate based on new data --- is naturally performed using 
Bayes's theorem (e.g.\ Sivia 1996; Jaynes 2003).   

The layout of the paper is as follows. In \S\,2 a simple approach
to flare prediction using only the past record of flaring from an active
region [previously presented in Wheatland (2001)] is reiterated. 
In \S\,3 the new method of prediction, 
combining existing methods and information from observed flare statistics,
is described.
In \S\,4 simulations are presented showing how the method uses the
observed flaring record, and in \S\,5 the results are discussed.

\section{Wheatland (2001)}

Wheatland (2001) presented a method for flare prediction using
only observed flare statistics and the assumptions that flares obey
Poisson statistics in time, and power-law statistics in size,
elaborating on a suggestion by Moon et al.\ (2001). 
The approach is briefly reiterated here, since it is part of the new
method. 

First assume that there is a threshold size $S_1$ above which
all events occurring in an active region are observed, so that the
distribution~(\ref{eq:pldist}) applies for events above that size. 
The total rate of events larger than $S_1$ is then
\begin{equation}
\lambda_1=\int_{S_1}^{\infty}N(S)dS=A(\gamma -1)^{-1}
  S_1^{-\gamma+1},
\end{equation} 
assuming $\gamma>1$. Hence the frequency-size distribution may be 
rewritten
\begin{equation}\label{eq:fdist}
N(S)=\lambda_1(\gamma-1)S_1^{\gamma-1}S^{-\gamma}.
\end{equation}
Suppose the probability of a big event in a given period $\Delta T$ is 
required, where by big we mean an event at least as large as 
$S_2$. According to the distribution~(\ref{eq:fdist})
the rate of events larger than $S_2$ is
\begin{equation}\label{eq:rate_big} 
\lambda_2=\lambda_1
  \left( \frac{S_1}{S_2}\right)^{\gamma-1}. 
\end{equation}

Applying the Poisson model of flare occurrence, the probability of at 
least one big event during a period $\Delta T$ is given by Poisson 
statistics as
\begin{equation}\label{eq:prob_big}
\epsilon =1-\exp(-\lambda_2 \Delta T).
\end{equation}

Equations~(\ref{eq:rate_big}) and~(\ref{eq:prob_big}) provide the
required estimate. The quantities $S_1$, $S_2$ and $\Delta T$ are chosen, 
and then the parameters $\lambda_1$ and $\gamma$ (if the precise value
of $\gamma$ is assumed unknown) need to be 
estimated from the past history of flaring of the active region.
Wheatland (2001) assumed that $\gamma$ is the same for all active 
regions, and hence known (see Wheatland 2000), 
and estimated $\lambda_1$ using the
Bayesian procedure of Scargle (1998).

The rationale behind the method of Wheatland (2001) is that the 
flare frequency-size distribution is steep so there are very many small 
events, which allows $\lambda_1$ to be estimated relatively accurately 
from the observed history of flaring in an active region. Hence the
estimate of $\epsilon$ should be relatively accurate. To make this
point quantitative, note that from Equations~(\ref{eq:rate_big}) 
and~(\ref{eq:prob_big}) the uncertainty in the estimate of the
probability $\epsilon $ is given approximately by
\begin{equation}
\frac{\sigma_{\epsilon}}{\cal \epsilon}
  =\frac{\lambda_1 \Delta T (S_1/S_2)^{\gamma-1}}
  {\exp[\lambda_1 \Delta T (S_1/S_2)^{\gamma-1}] -1}
  \frac{\sigma_1}{\lambda_1},
\end{equation}
where $\sigma_1$ is the uncertainty in $\lambda_1$, and where we have 
ignored any uncertainty in $\gamma$. Assuming $S_2\gg S_1$ leads to 
$\sigma_{\epsilon}/\epsilon \approx \sigma_1/\lambda_1$.
If the rate $\lambda_1$ is determined from 
$M$ observed events, then for Poisson statistics we expect
$\sigma_1/\lambda_1=M^{-1/2}$, and hence 
\begin{equation}\label{eq:unc}
\frac{\sigma_{\epsilon}}{\epsilon}\approx M^{-1/2}.
\end{equation}
Equation~(\ref{eq:unc}) provides a crude estimate of the accuracy of the
method. To achieve a 10\% accuracy in the estimate requires of order
100 observed events.

\section{New method}

\subsection{Approach}

The Wheatland (2001) method shows how to use the flaring record
for an active region to make a flare prediction, but it ignores the 
other information which is normally the basis of prediction. It is 
sensible to combine all of the available information, and in this 
section we consider how to do this.

We assume that a sequence of events with sizes $s_1,s_2,...,s_M$ 
(all larger than $S_1$) are observed to occur at times 
$t_1< t_2< ...< t_M$ respectively in an active region. 
These events occur within an observing interval which starts at 
time $t_{\rm sta}$ and ends at time $t_{\rm end}$. We also have
additional information, which we label $I$, including our 
knowledge of the phenomenological rules of flare statistics, and
e.g.\ the McIntosh classification of the active region. 
The problem is then to estimate $\epsilon$, the probability of a big
event, based on the data and the additional information $I$. 
By `estimating $\epsilon$' we strictly mean that we want to calculate 
a probability distribution for the quantity $\epsilon$, based on the 
available information. The peak of this distribution
is our most likely value for the probability of occurrence of a big
flare, and the width of the distribution is a measure of the
uncertainty of that value. To do this we proceed as follows. 
First we estimate (calculate probability distributions for) 
$\lambda_1$ and $\gamma$ based on the available information, and then 
we use these distributions to estimate $\lambda_2$. Then we use this 
distribution together with the relationship~(\ref{eq:prob_big}) to 
estimate the desired quantity $\epsilon$. We now consider each of 
these steps in turn.

\subsection{Estimating $\gamma$}

First we consider the calculation of 
$P_{\gamma}(\gamma )$, the probability distribution for 
the power-law index 
$\gamma$.\footnote{In the following probability distributions are given
labels such as $P_{\gamma}(\gamma)$ when the actual functional form 
of the distribution is needed. When this is not the case 
the generic label ${\rm prob}(...)$ is used to denote a 
distribution.} 
As mentioned in the Introduction, 
Wheatland (2000) found that the index $\gamma$ 
is independent of active region for a set of hard X-ray events, 
although the statistics underlying 
the study were somewhat poor. If $\gamma$ is the same in all active 
regions then the 
observations $s_1,s_2,...,s_M$ can be replaced by a larger set of 
events over many active regions. We return to this point in \S\,3.4, 
but for now admit the possibility that $\gamma$ is different in different 
active regions, and consider its estimation based on data for the given 
active region alone. 

Bai (1993) has shown how to estimate a power-law index for a set of 
data, using `maximum likelihood'. Following Bai, the likelihood 
function, that is the probability of the observed data 
$D=\{s_1,s_2,...,s_M\}$ given the model, is (assuming $\gamma>1$)
\begin{equation}\label{eq:gam_like}
{\rm prob}(D | \gamma, I )
  \propto \prod_{i=1}^{M}(\gamma-1)(s_i/S_1)^{-\gamma},
\end{equation}
where $I$ stands for all additional information, including knowledge of 
the phenomenological rule~(\ref{eq:pldist}). We note that this
expression requires $\gamma >1$, which follows from the requirement that
the probability distribution for size $S$ is normalized over all $S$ larger
than $S_1$. It is not necessary to introduce an upper cutoff for $S$ in 
the present treatment (provided $\gamma >1$), although an upper cutoff 
is necessary to ensure that the mean flare size is finite, if 
$\gamma<2$. We will return to this point in \S\,5.

Bayes's theorem may be used to convert the likelihood into the 
probability of the model given the data, which is what we are 
interested in: 
\begin{equation}\label{eq:p_gam_bayes}
{\rm prob}(\gamma | D,I)
\propto 
  {\rm prob}(D | \gamma,I)\times {\rm prob}(\gamma,I ),
\end{equation}
where ${\rm prob}(\gamma,I )$ is the `prior distribution' for
$\gamma$, i.e.\ the distribution we would assign to $\gamma$ in
the absence of the data (e.g.\ Sivia 1996). A choice needs to 
be made for this distribution, and a common choice is to assume 
a constant value within minimum and maximum values $\gamma_1$ and
$\gamma_2$ respectively:
\begin{equation}
{\rm prob} (\gamma |D,I) = \left\{
  \begin{array}{ll}
  (\gamma_2-\gamma_1)^{-1} & \mbox{if $\gamma_1\leq \gamma \leq
\gamma_2$}
  \\
  0 & \mbox{else,}
\end{array}
\right.
\end{equation}
which is referred to as a `uniform prior'.
We note that for a uniform prior the most likely value of 
$\gamma$ is the maximum of the likelihood function:
\begin{equation}\label{eq:gam_ML}
\gamma^{\ast}=\frac{M}{\sum_{i=1}^{M}\ln (s_i/S_1)}+1,
\end{equation}
which is the maximum likelihood estimate of $\gamma$ found by Bai. 

We can identify ${\rm prob} (\gamma | D,I)$ with 
$P_{\gamma}(\gamma)$, and then Equations~(\ref{eq:gam_like}) 
and~(\ref{eq:p_gam_bayes}) 
give the required `posterior distribution' for $\gamma$:
\begin{equation}\label{eq:prob_gam}
P_{\gamma}(\gamma)= C \frac{(\gamma-1)^{M}}{\pi^{\gamma}}\Gamma (\gamma),
\end{equation}
where
\begin{equation}
\pi=\prod_{i=1}^M\frac{s_i}{S_1},
\end{equation}
and where we have relabelled the prior distribution $\Gamma (\gamma)$.
The normalizing factor $C$ is determined by the requirement
$\int_{1}^{\infty}P_{\gamma}(\gamma)d\gamma=1$.\footnote{In the 
following all normalizing factors are labelled $C$, although they 
refer to different values. It is understood that in each case the 
value $C$ is to be determined by integration.} For a uniform prior
the integral may be performed, leading to
\begin{equation}
C=\frac{(\gamma_2-\gamma_1) \pi (\ln \pi )^{M+1}/M!}
  {P[M+1,(\gamma_2-1)\ln\pi ]
  - P[M+1, (\gamma_1-1)\ln \pi ]},
\end{equation}
where $P (a,x)$ denotes the incomplete Gamma function~\cite{abr&ste64}.

Before proceeding we present a rough estimate of the uncertainty in 
the most likely value of $\gamma$ based on the distribution 
$P_{\gamma}(\gamma)$ with a uniform prior. 
Assuming Gaussian behavior in the vicinity of
the peak, the width of the distribution~(\ref{eq:prob_gam}) is 
$\sigma_{\gamma}\approx [L^{\prime\prime}(\gamma^{\ast})]^{-1/2}$, where 
$L(\gamma)=-\ln P_{\gamma}(\gamma)$, and where $\gamma^{\ast}$ is the
location of the peak of the distribution (Sivia 1996). This leads to
$\sigma_{\gamma}\approx M^{1/2}/\ln\pi$, and using 
Equation~(\ref{eq:gam_ML}) gives
\begin{equation}\label{eq:sig_gam}
\sigma_{\gamma}\approx (\gamma^{\ast}-1)M^{-1/2}.
\end{equation}

\subsection{Estimating $\lambda_1$}

Next we consider the calculation of $P_1(\lambda_1)$, the distribution 
of the rate $\lambda_1$ of flares larger than $S_1$.
This is a more difficult problem because the rate of flaring in an active 
region may vary with time~(see e.g.\ Wheatland 2001). However,
observations suggest that a piecewise-constant Poisson process
provides a good model for the way flares occur in time in 
individual active regions. 

We assume that a period of time of duration $T^{\prime}\leq T$ immediately 
prior to $t_{\rm end}$ is identified (i.e.\ from $t=t_{\rm end}-T^{\prime}$
to $t=t_{\rm end}$) during which time flare occurrence is consistent
with a constant-rate Poisson process. 

One approach to identifying the necessary period of time has been 
presented by Scargle (1998), who showed how to select a piecewise-constant 
Poisson model to describe an observed sequence of events. When applied 
to a sequence of events at times $t_1< t_2< ... < t_M$ the Scargle method 
gives a sequence of times $t_{B\it 0}< t_{B1}<...<t_{BK}$ 
at which the rate is determined to change 
(where $t_{B0}=t_{\rm sta}$ and $t_{BK}=t_{\rm end}$ are the start and
end of the observing period), and a corresponding sequence 
$\lambda_{B1},\lambda_{B2},...,\lambda_{BK}$ of rates. The sequence 
of times and rates is called a set of  `Bayesian blocks'. In this
case we identify $T^{\prime}$ with $t_{BK}-t_{B(K-1)}$.
We note that the original Bayesian blocks procedure [which was used 
e.g.\ by Wheatland (2001)] does not necessarily select the best 
piecewise-constant model. Recently Scargle has found a computationally 
feasible way to determine the optimal decomposition (Scargle, private 
communication, 2003). We begin by assuming this method (or another 
method) has been applied to the data, to determine the required period 
$T^{\prime}$ prior to the end of observations.

A probability distribution for the rate $\lambda_1$ is then be
determined as follows. We assume that $M^{\prime}\leq M$ events are observed 
during the selected period $T^{\prime}$. The probability of the observed 
data $D^{\prime}$ (strictly this comprises not just the number of events 
but also their times) given a Poisson model with rate $\lambda_1$ is 
\begin{equation}\label{eq:pdkmk}
{\rm prob} (D^{\prime}|\lambda_1,I)\propto \lambda_1^{M^{\prime}}
  e^{-\lambda_1T^{\prime}},
\end{equation}
where we retain only the dependence on $\lambda_1$ on the
right hand side of this equation, and where we formally recognise any 
additional information by the dependence on $I$.
Bayes's theorem may be used to turn this likelihood into a probability 
of the model given the data, and the additional information: 
\begin{equation}\label{eq:pmkdk}
{\rm prob}(\lambda_1|D^{\prime},I)\propto 
  {\rm prob}(D^{\prime}|\lambda_1,I)\times {\rm prob} (\lambda_1,I),
\end{equation}
where ${\rm prob}(\lambda_1,I)$ is the prior distribution for the rate. 

The prior distribution ${\rm prob} (\lambda_1,I)$ represents the 
estimate of the rate of flaring for the active region in the absence 
of any data. This distribution allows the incorporation of any additional
information we have about the expected rate of flaring, not including 
the actual data. To make this concrete, we will consider the case that
the additional information is the McIntosh classification of the sunspots
associated with the active region, although we stress that any other
additional information can also be incorporated.
When the additional information is the McIntosh classification, 
a suitable prior distribution can be 
constructed from historical records of the observed rates of events 
above size $S_1$ for every active region of the same class. 
This is a generalization of the analysis underlying present flare 
prediction methods based on McIntosh classification, which considers 
only the mean flaring rate extracted from historical data. Hence we 
propose the construction of distributions of flaring rate for each 
McIntosh classification. We assume these are available, and label the 
appropriate distribution 
$\Lambda_{\rm MC} (\lambda_1)$, where MC denotes McIntosh 
classification. Equation~(\ref{eq:pmkdk}) then becomes
\begin{equation}\label{eq:prob_lam1}
P_1(\lambda_1)=C\lambda_1^{M^{\prime}} e^{-\lambda_1T^{\prime}}
  \Lambda_{\rm MC} (\lambda_1),
\end{equation}
where we have identified ${\rm prob}(\lambda_1|D^{\prime},I)$ with
$P_1(\lambda_1)$, and and where $C$ is the normalization factor. This
is the required posterior distribution for $\lambda_1$.

It should be noted that the distribution~(\ref{eq:prob_lam1}) explicitly
uses only a subset of all flares observed in an active region, 
i.e.\ the $M^{\prime}\leq M$ flares observed during the interval 
$T^{\prime}\leq T$. Previous 
data contribute only to the determination of the interval $T^{\prime}$. The
motivation is that when the rate changes, the old rate is no
longer relevant for future prediction. For many active regions the
observed rate appears to be constant during a transit of the disk, or
at least no rate change is detectable (e.g.\ Wheatland 2001), in which
case all observed flares contribute explicitly to the inference.

Before proceeding we note two simple results for 
Equation~(\ref{eq:prob_lam1}) with a uniform prior.
First, it is easy to see that with a uniform prior the maximum of this 
distribution occurs at $M^{\prime}/T^{\prime}$.
Second we note the well known result that for large $\lambda_1T^{\prime}$ 
and neglecting the prior, Equation~(\ref{eq:prob_lam1}) 
approximates a Gaussian with a width 
\begin{equation}\label{eq:sig_lam}
\sigma_1\approx \frac{(M^{\prime})^{1/2}}{T^{\prime}},
\end{equation} 
which is consistent with the arguments at the end of \S\,2. 

\subsection{Estimating $\epsilon$}

The probability distribution $P_2(\lambda_2)$ for the rate $\lambda_2$
of flares larger than $S_2$ may be constructed from the distributions 
$P_1(\lambda_1)$ and $P_{\gamma}(\gamma)$ using 
Equation~(\ref{eq:rate_big}). Specifically we have 
$\lambda_2=\lambda_1(S_1/S_2)^{\gamma-1}$, 
and hence
\begin{equation}
P_2(\lambda_2)= 
  \int_1^{\infty}d\gamma \int_0^{\infty} d\lambda_1 P_1(\lambda_1)
  P_{\gamma}(\gamma)\delta
  \left[ \lambda_2-\lambda_1(S_1/S_2)^{\gamma-1}\right],
\end{equation} 
and performing the integral over $\lambda_1$ leads to
\begin{equation}\label{eq:P2}
P_2(\lambda_2) = 
  \int_1^{\infty} 
  d\gamma P_{\gamma}(\gamma)\left(\frac{S_2}{S_1}\right)^{\gamma-1} 
  P_1\left[\lambda_2 \left(\frac{S_2}{S_1}\right)^{\gamma-1} \right].
\end{equation}

The quantity we are interested in is $\epsilon$, the probability of
an event bigger than $S_2$ occurring in an interval $\Delta T$. 
The probability distribution $P_{\epsilon}(\epsilon)$ for this
quantity may be contructed from the distribution for $\lambda_2$ by 
a change of variable. 
Specifically, from Equation~(\ref{eq:prob_big}) we have 
$\lambda_2=-\ln (1-\epsilon)/\Delta T$, and hence  
\begin{eqnarray}\label{eq:prob_pbig}
P_{\epsilon}(\epsilon)&=&P_2\left[\lambda_2(\epsilon )\right]
\left|\frac{d\lambda_2}{d\epsilon }\right| \nonumber \\
&=& P_2\left[-\frac{\ln (1-\epsilon )}{\Delta T} \right]
  \frac{1}{\Delta T (1-\epsilon ) }.
\end{eqnarray}
Using Equations~(\ref{eq:prob_gam}), (\ref{eq:prob_lam1}), and
(\ref{eq:P2}) in~(\ref{eq:prob_pbig}) leads to
\begin{equation}\label{eq:pbig_general}
P_{\epsilon}(\epsilon )=
  \int_1^{\infty} d\gamma \, f(\epsilon,\gamma),
\end{equation}
where
\begin{eqnarray}\label{eq:fjoint}
f(\epsilon,\gamma)&=&C\left[-\ln (1-\epsilon )\right]^{M^{\prime}} 
  (\gamma-1)^M\Gamma (\gamma ) 
  \left[\frac{(S_2/S_1)^{M^{\prime}+1}}{\pi}\right]^{\gamma}
  \nonumber \\ 
  &\times& (1-\epsilon )^{\left(T^{\prime}/\Delta T\right)
    \left(S_2/S_1\right)^{\gamma-1}-1}
  \Lambda_{\rm MC} \left[-\frac{\ln (1-\epsilon )}{\Delta T} 
  \left(\frac{S_2}{S_1}\right)^{\gamma-1} \right]
\end{eqnarray}
is the joint probability 
distribution for $\epsilon$ and $\gamma$. The normalization factor
$C$ is obtained by requiring that 
$\int_{0}^{1}P_{\epsilon}(\epsilon)d\epsilon=1$. We note that
$P_{\gamma}(\gamma)$ and $P_{\epsilon}(\epsilon)$ may be considered
to be marginal distributions of $f(\epsilon,\gamma)$ (i.e.\ they are
obtained by integration over $\epsilon$ and $\gamma$ respectively). 
However, Equation~(\ref{eq:prob_gam}) gives the distribution for
$\gamma$ directly.

As noted in \S\,3.2, observations suggest that $\gamma$ is the same
in all active regions, in which case the index can be determined very 
accurately from events over many active regions using 
Equation~(\ref{eq:gam_ML}). If the estimate is $\gamma^{\ast}$, 
then we can consider the prior distribution for $\gamma$ to be
$\Gamma (\gamma) = \delta (\gamma-\gamma^{\ast})$, and 
Equation~(\ref{eq:pbig_general}) simplifies to 
\begin{equation}\label{eq:pbig_simp}
P_{\epsilon}(\epsilon ) =
  C\left[-\ln (1-\epsilon ) \right]^{M^{\prime}} 
  (1-\epsilon )^{\left( T^{\prime}/\Delta T\right)
  \left(S_2/S_1\right)^{\gamma^{\ast}-1}-1} 
  \Lambda_{\rm MC} \left[-\frac{\ln (1-\epsilon )}{\Delta T} 
  \left(\frac{S_2}{S_1}\right)^{\gamma^{\ast}-1} \right].
\end{equation}

Equations~(\ref{eq:pbig_general}), (\ref{eq:fjoint})
and~(\ref{eq:pbig_simp}) are the required expressions for the posterior 
probability distribution for $\epsilon$.

\section{Simulations}

We present two simulations demonstrating the application of the
method to synthetic data. These simulations omit the inclusion of 
other information via the prior 
$\Lambda_{\rm MC} (\lambda_1)$, so they illustrate only how the
method performs using the observed data. 

First we consider the case that $\gamma$ is assumed to be known.
Ten days of flaring were simulated by producing a sequence of
event times as a Poisson process in time with a rate $\lambda_1=0.5$
per day for the first five days, and with a rate $\lambda_1=5.0$
per day for the second five days. Each event was assigned a size 
according to a power law distribution with an index $\gamma=1.8$, 
above the threshold size $S_1=1$ (in arbitrary units). Figure~1
illustrates a typical simulation. The first (upper) panel shows the 
size of each event versus the time at which the event occurred.
In this case there were 31 events. The simulation applies the method
to the problem of predicting the probability of a big event occurring 
during the next day ($\Delta T=1$ day) at the end of the ten days. 
The size of a big event was taken to be 
$S_2=100$. The original Bayesian blocks procedure~(Scargle 1998) was 
applied to the event time series to determine a decomposition into a 
sequence of piecewise-constant intervals and rates. The second panel 
of Figure~1 shows the result of this process: 
the solid lines indicate the rate as a function of time 
inferred by the Bayesian blocks procedure, and the dotted lines indicate 
the true rate versus time. The Bayesian blocks procedure correctly
identifies a two-rate model as the most likely model, and identifies 
the approximate time of the change in rate. The third panel shows the
probability distribution $P_{\epsilon}(\epsilon)$ obtained from 
Equation~(\ref{eq:pbig_simp}) with a uniform prior for $\lambda_1$, 
and with $M^{\prime}$ and $T^{\prime}$ equal to the number of events in the 
second Bayesian block and the duration of the second Bayesian block 
respectively. The dotted vertical line in this panel is the true value
of $\epsilon$. 
We see that, even for a relatively small number of events, the method is 
able to provide a good estimate of the probability of a big event. The
width of the inferred distribution for $\epsilon$ is consistent with
Equation~(\ref{eq:unc}).

\begin{figure}
\epsscale{0.7}
\plotone{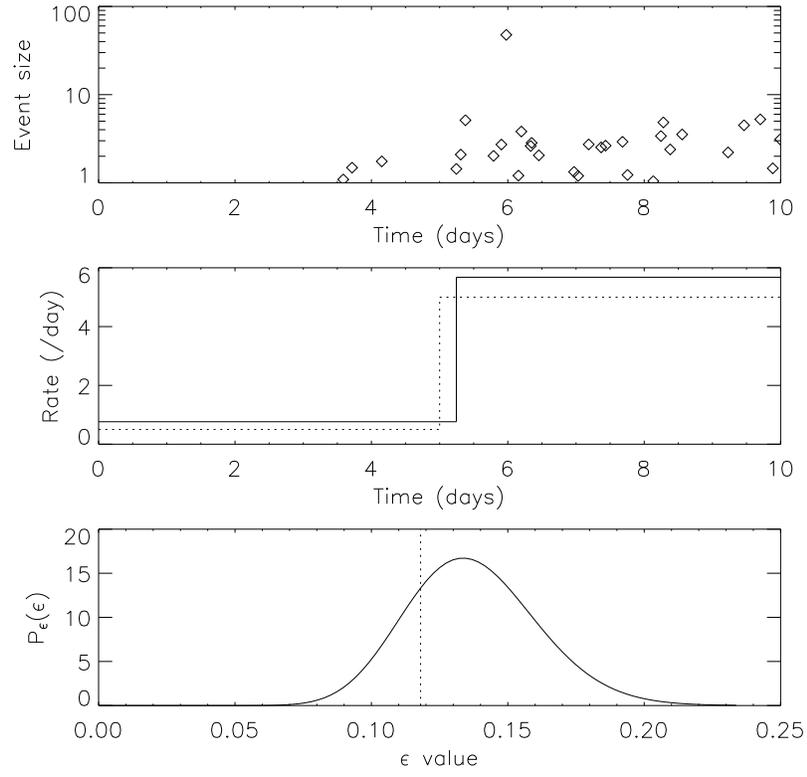}
\caption[f1.eps]{Simulation of 10 days of flaring and application of 
the prediction method, assuming $\gamma$ is known.}
\end{figure}

Second we consider the more difficult case of simultaneously
estimating $\gamma$ and $\lambda_1$. Ten days of flaring were again 
simulated, with a rate $\lambda_1=1$ per day for the first five days, 
and a rate $\lambda_1=10$ per day for the second five days. Larger 
rates were chosen to provide more events for the inference, but the 
other parameters were kept the same as in the first simulation. 
Figure~2 illustrates the results of a typical simulation. The 
first (upper) panel shows the time history of events --- in this case
57 events occurred. The second panel shows the result of a Bayesian
blocks decomposition of the data (solid lines) together with the
true rate versus time (dotted lines). Once again the Bayesian blocks
procedure correctly identifies a two-rate model as the most likely
model, and identifies the approximate time of the change in rate.
The third panel shows  the result of using Equation~(\ref{eq:prob_gam})
--- with a uniform prior with $\gamma_1=1.25$ and $\gamma_2=2.25$ ---
to construct the distribution for $\gamma$. The dotted vertical line in this 
panel shows the true value of $\gamma$.
The fourth panel of Figure~2 shows the distribution for $\epsilon$
constructed using Equation~(\ref{eq:pbig_general}), with 
$M=57$, with $M^{\prime}$ and $T^{\prime}$ obtained from the second 
Bayesian block, and with uniform prior distributions for $\gamma$ and 
$\lambda_1$. 
The dotted vertical line indicates the true value. From this simulation 
we see that a reasonable estimate for $\epsilon $ is obtained for a 
relatively small number of events. 

\begin{figure}
\epsscale{0.7}
\plotone{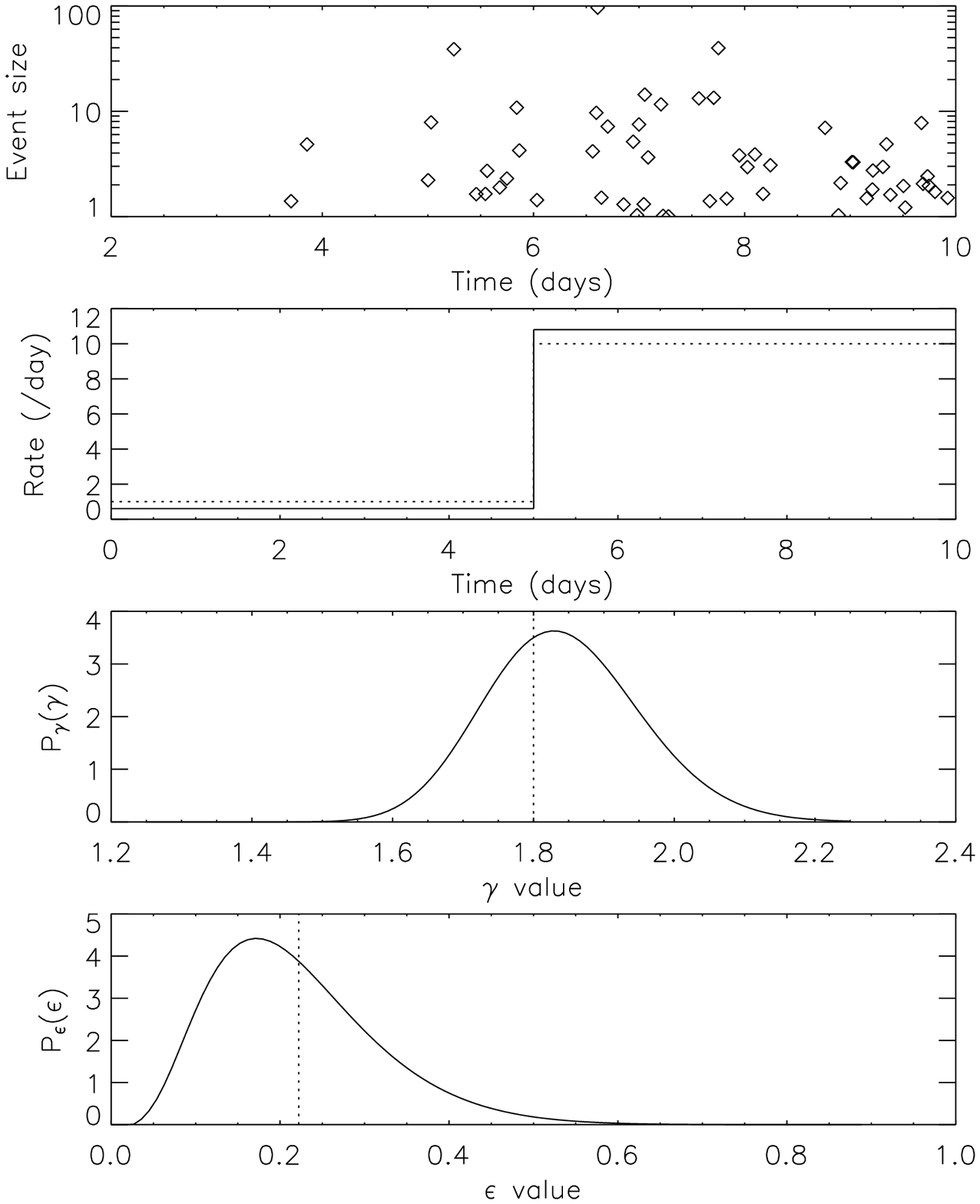}
\caption[f2.eps]{Simulation of 10 days of flaring and application of the
prediction method, assuming $\gamma$ is unknown.}
\end{figure}

The distribution for $\epsilon $ obtained in the lower panel of
Figure~2 is quite broad.
A basic reason is that $\epsilon$ depends sensitively on $\gamma$ 
because of its appearance as an exponent in
Equation~(\ref{eq:rate_big}), and $\gamma$ has a range of possible
values, as shown in the third panel of Figure~2.
This effect may be seen by considering
$f(\epsilon,\gamma)$ [defined by Equation~(\ref{eq:fjoint})],
which is the joint distribution of $\epsilon$ and $\gamma$. Figure~3
shows a contour plot of $f(\epsilon,\gamma)$ for the simulation depicted
in Figure~2. The dotted vertical and horizontal lines are the true values
of $\epsilon$ and $\gamma$ respectively.
The dashed curve is defined by 
$\epsilon=1-\exp[-(M^{\prime}/T^{\prime})(S_1/S_2)^{\gamma-1}\Delta T ]$, 
and the contours of $f(\epsilon,\gamma)$ are observed to be stretched
out along this curve. The practical implication of this figure is that
accurate estimation of $\epsilon$ depends on accurate estimation
of $\gamma$. In practice $\gamma$ is known a priori quite accurately,
but in this simulation we have assumed that $\gamma$ is initially unknown
(within the range 1.25 to 2.25), to illustrate the process of
inference. 
 
\begin{figure}
\epsscale{0.7}
\plotone{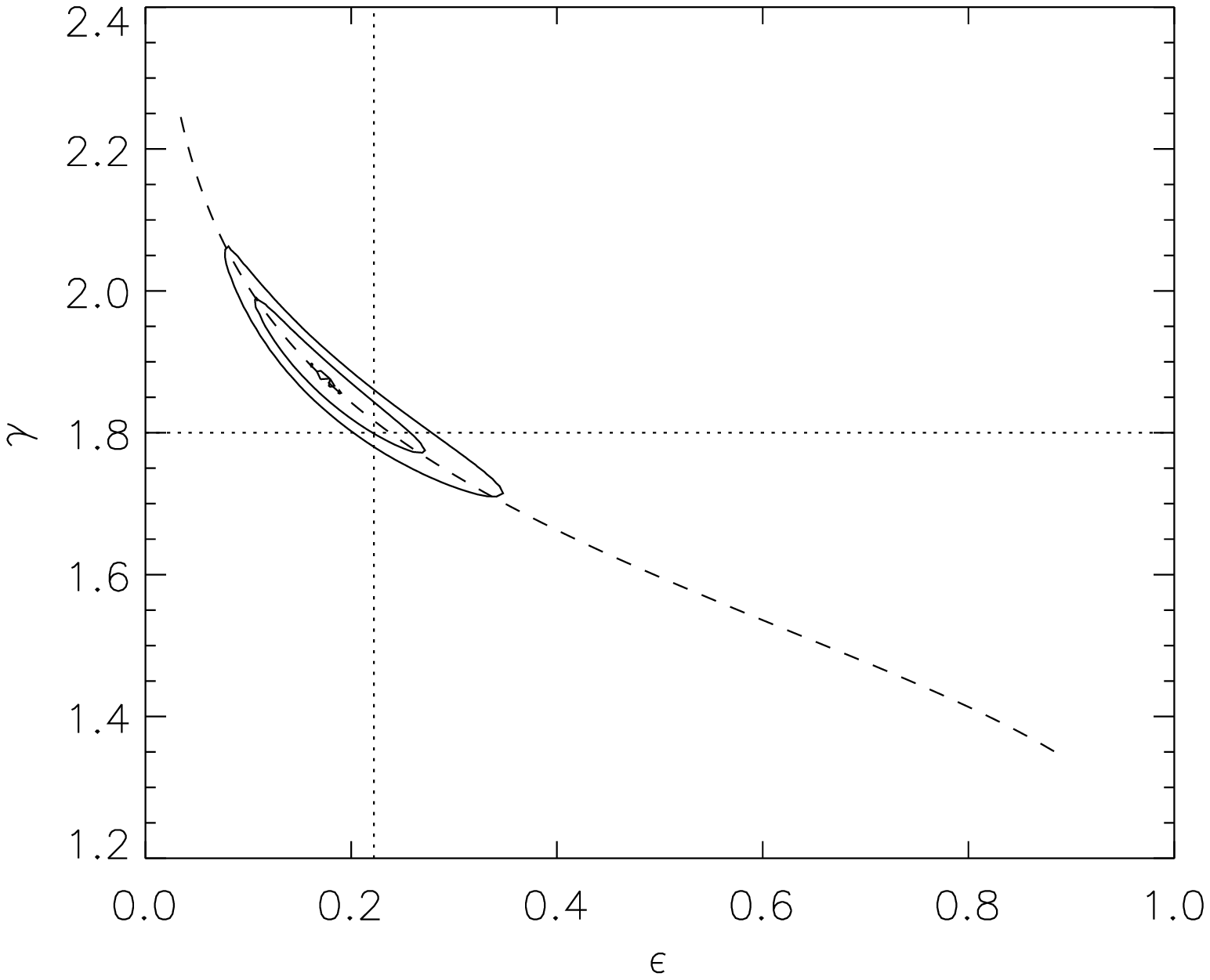}
\caption[f3.eps]{Contour map of the joint probability of $\epsilon$ and
$\gamma$, for the simulation in Fig.~2.}
\end{figure}

\section{Discussion}

Existing methods of solar flare prediction do not make complete
use of an important source of information: the time history of flares
already observed in the active region of interest, in particular 
frequently occurring small events.
A new method for flare prediction is presented which exploits the 
observed history of flaring from an active region to improve an initial
prediction, which e.g.\ may come from one of the existing methods. 
To make the example concrete we may think of the initial prediction
coming from from the McIntosh sunspot classification, which is a common 
basis for prediction. This background information provides an initial 
estimate for the expected flaring rate through a prior distribution 
$\Lambda_{\rm MC}(\lambda_1)$, which represents the probability that
the flaring rate above a (small) size $S_1$ is $\lambda_1$, given
historical rates of occurrence of flares for the given McIntosh 
class. Bayes's theorem is then used to estimate the probability 
$\epsilon$ of observing a large flare (above size $S_2$) in a given 
period of time, based on this prior information and on the sequence of 
flares already produced by the active region, and assuming simple
phenomenological rules describing the occurrence of flares. 
In this paper the basic theory behind the inference of $\epsilon$ 
based on observed data is presented. The inclusion of background
information [i.e.\ the construction of the priors 
$\Lambda_{\rm MC}(\lambda_1)$] is yet to be done.

The method relies on event sizes following the phenomenological 
law~(\ref{eq:pldist}). Some studies of very small extreme 
ultraviolet events (`nanoflares') suggest that their thermal energies 
follow a steeper distribution than energies of large events 
(e.g.\ Krucker and Benz 1998; Parnell and Jupp 2000), although this 
remains controversial (e.g.\ Aschwanden and Parnell 2002). 
From the point of view of the prediction method presented here,
the uncertainty over the low-size end of the distribution is irrelevant 
provided events significantly larger than nanoflares are used.
In any case the observed distributions from many active 
regions may be examined as a check on Equation~(\ref{eq:pldist}).
A related point is that the distribution~(\ref{eq:pldist}) requires 
a cutoff at large sizes on energetics grounds, and neglect of this
cutoff will lead to the number of large flares being overestimated. 
A cutoff will be incorporated before the method is applied to real data.

The choice of the quantity $S$ has not been addressed, although a good
choice is likely to be important to the method. Most flare forecasting
deals with soft X-ray events, in particular prediction of GOES
(Geostationary Observational Environmental Satellite) M and X class 
events (events with peak fluxes greater than $10^{-5}$W/m$^2$
and $10^{-4}$W/m$^2$ respectively in the 1-8 Angstrom band observed by
the satellites). A practical motivation for this is that flare
soft X-ray emission causes disturbances of the ionosphere which affect 
shortwave radio communication, and there is a need to predict these
occurrences. A disadvantage of using GOES events is that they are not 
ideal for flare statistics e.g.\ because of problems with event selection 
due to the large background in soft X-ray (see Wheatland 2001). 

A number of other issues also need to be considered before the method is
implemented with real data. A point neglected so far is that active regions 
evolve, so that predictions based on the traditional methods also 
change with time. For example, an active region evolves through McIntosh 
classifications (e.g.\ Bornmann, Kalmbach, Kulhanek, and Casale 1990). 
Changes in background information such as this should be incorporated 
through changes in the prior, and this question will be considered in more 
detail in future work. A related point concerns the construction of the 
prior distributions for rate. It is likely that the McIntosh classification 
will be used, although other possibilities will be considered. The
problem is then to determine the probability of a given McIntosh class 
having a given rate, based on observed flaring sequences in the
historical record for active regions of that class.
The details of this calculation will be addressed in future work. 
 
Finally, as with all methods of forecasting, it is essential to 
test the reliability of the method. It is straightforward to compare, 
after the fact, the number of predicted and the number of observed 
events for a large sample of active regions. The method presented here 
will be implemented and tested in this way, and the results compared 
with existing methods of prediction.

\section*{Acknowledgements}

M.S.W. acknowledges the support of an Australian Research Council
QEII Fellowship, and thanks Richard Thompson and Garth Patterson
of the Ionospheric Prediction Service for useful discussions. The
comments of an anonymous referee have also helped to improve the 
paper.

\end{document}